\documentclass[twocolumn,twoside]{revtex4}
\usepackage{graphicx}
\usepackage{fancyhdr}
\def\be{\begin{equation}}
\def\ee{\end{equation}}
\def\bea{\begin{eqnarray}}
\def\eea{\end{eqnarray}}

\begin{document}
% FTPI-MINN-19/16, UMN-TH-3825/19

\title{Deciphering the XYZ States}


\author{M. B. Voloshin}
\affiliation{William I. Fine Theoretical Physics Institute, University of
Minnesota, Minneapolis, MN 55455, USA \\
School of Physics and Astronomy, University of Minnesota, Minneapolis, MN 55455, USA} %\\ and \\
\affiliation{Institute of Theoretical and Experimental Physics, Moscow, 117218, Russia}

\begin{abstract}
I give a brief account of current topics in description of exotic multiquark mesonic resonances with hidden heavy ($c$ or $b$) flavor, the so-called XYZ states, in terms of hadronic molecules and hadro-quarkonium systems.  Also discussed are the recently observed hidden-charm pentaquarks including additional ways of producing them in experiments.
\end{abstract}

\maketitle

\thispagestyle{fancy}


\section{Introduction}
Experiment has revealed a number of states with hidden charm and hidden bottom that do not fit the standard quark model template of mesons consisting of a quark and antiquark and baryons consisting of three quarks. Thus far about a dozen or more mesons above the open charm threshold require presence inside them of a light quark-antiquark pair along with a hidden charm $c \bar c$ quark pair. Two isotopic doublets of mesons with a heavier $b \bar b$ quark pair with a similar four-quark structure have been observed. More recently, sightings of pentaquarks consisting of three light quarks and a $c \bar c$ pair have been reported.  This family of hadronic states, referred to as exotic, and expanding from the first observation~\cite{bellex} of the charmonium-like peak $X(3872)$ up to the most recently reported~\cite{lhcbpc3} narrow pentaquarks,  conspicuously challenges the theory for explaining their `internal workings' and  possibly predicting their yet unknown properties and new states of similar nature. 

\begin{figure}[h]
\centering
\includegraphics[width=80mm]{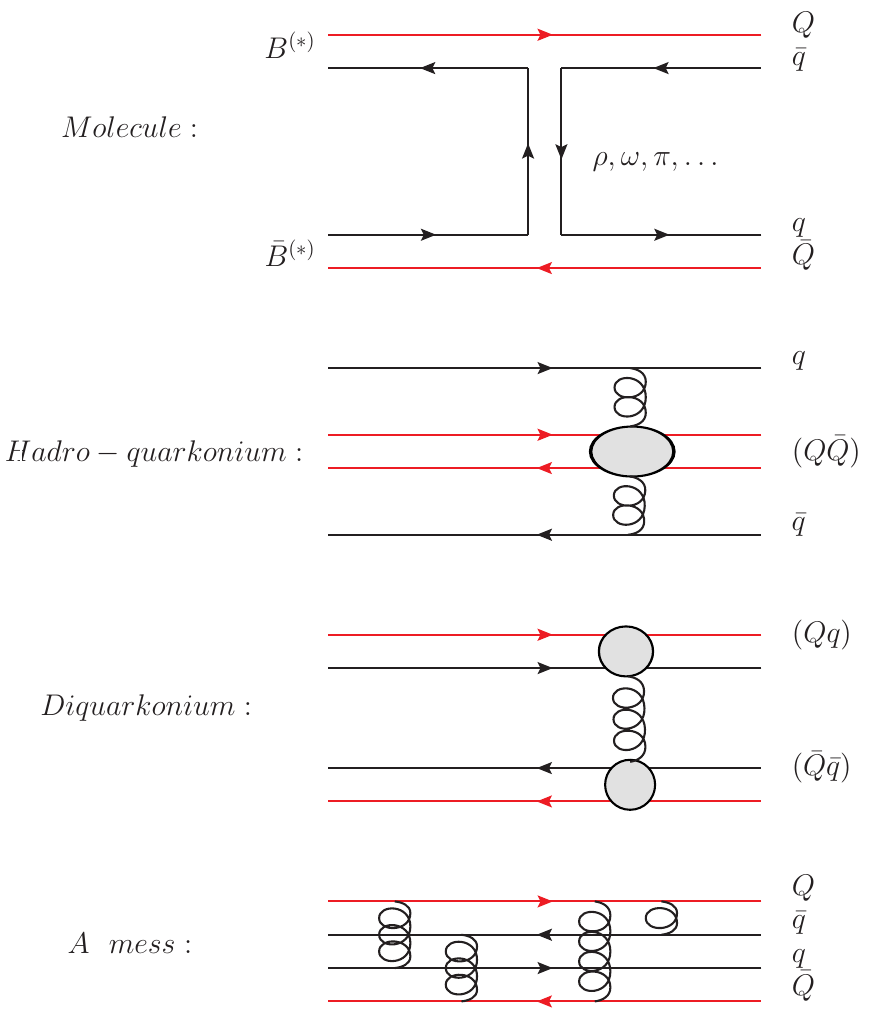}
\caption{Types of internal dynamics of exotic heavy mesons.} \label{inside}
\end{figure}

Detailed reviews of the experimental status of the new states and of various theoretical models can be found in a number of very recent papers ~\cite{dsz,Guo17,Ali17,Liu}. Here I present a brief account of the theoretical situation as I see it. The existing schemes for description of the exotic states are based on the idea  that the complicated multiquark dynamics splits into simpler few-body correlations that may be tractable theoretically. The discussed types of such two-body correlations are shown in  Figure~\ref{inside}. Although all of the shown configurations are likely present, to an extent, in the XYZ mesons, as will be discussed, there are good reasons to believe (or at least a strong hope) that the configurations of two first types, molecules and hadro-quarkonium, are dominant in some of the observed exotic states. On the contrary, it can be argued that the third structure in Figure~\ref{inside} corresponding to significant correlations within diquarks and antidiquarks does not have a justification within QCD dynamics. As to the most theoretically `unpleasant' fourth structure, where all interactions are of similar strength and no few-body correlations can be considered as more important than other, one can only hope that it is not very important in at least some of the exotic hadrons.
The study of baryonic exotic states, the pentaquarks, is still at an early stage, so that only preliminary remarks can be made at present regarding the internal dynamics of these baryons. 

In what follows I discuss the configurations of Figure~\ref{inside} within specific mesonic XYZ states, and then briefly discuss the hidden-charm pentaquarks.

\section{Molecules}

The notion of hadronic molecules made of heavy-light hadrons goes back to the following simple consideration~\cite{ov76}. The interaction between such hadrons, mediated by exchange of light quarks and antiquarks does not depend on the mass of the heavy quark in the limit, where that mass is large. On the other hand the kinetic energy is inversely proportional to the heavy mass. Thus for the states of heavy hadron pairs, where the light-quark exchange results in an attraction there should be bound or/and resonance levels for a sufficiently heavy heavy quark. In such a near threshold state, the motion of the constituents proceeds at long distances (much like the motion of the proton and neutron in deuteron), where the heavy hadrons largely retain their individual structure. The relevant size of such states can be estimated in terms of the binding/excitation energy $\delta$ (relative to the threshold) and the mass $M$ of the constituents as 
\bea
\label{rest}
&&r \sim 1/\sqrt{M \, |\delta|} \approx \\ \nonumber && \left \{ \begin{array}{c}
4.5\,{\rm fm} \sqrt{1\,{\rm MeV} \over |\delta|}~~~ { \rm charmonium-like}\\
2.8\,{\rm fm} \sqrt{1\,{\rm MeV} \over |\delta|} ~~~ { \rm bottomonium-like}~,
\end{array} \right .
\eea
so that the notion of molecules applies in a narrow mass band around the threshold with $\delta$ not exceeding 10-20\, MeV for charmonium-like states and not exceeding several MeV for the bottomonium-like ones.

It was however not known a priori whether charmed or bottom quarks would be `sufficiently heavy' until in 2003 the Belle experiment reported~\cite{bellex} observation of a charmonium-like peak $X(3872)$ in the $J/\psi \pi^+ \pi^-$ channel. The peak is very narrow~\cite{pdg}, $\Gamma < 1.2\,$MeV, and the mass of the peak practically coincides with the threshold for $D^{*0} \bar D^0$ charmed meson pair: $M(X) - M(D^{*0})- M(D^0) = -0.01 \pm 0.18\,$MeV. According to the present understanding (see e.g. in Ref.~\cite{Guo17}) the $X(3872)$ peak is dominantly a shallow bound or a virtual $J^{PC}= 1^{++}$ state of an $S$-wave pair of neutral charmed mesons $D^{*0} \bar D^0 + \bar D^{*0} D^0$. This structure successfully explains the isospin violation of order one that is evident from simultaneous existence (with a comparable rate) of the decays to the final states $J/\psi \rho^0$, $J/\psi \omega$~\cite{pdg,besxomega} and $\chi_{c1} \pi^0$~\cite{besxpi}.

A very clean example of molecular states is presented by the bottomonium-like $Z_b(10610)$ and $Z_b(10650)$ resonances~\cite{bellezb}, whose masses are within few MeV from respectively the $B^* \bar B$ and $B^* \bar B^*$ thresholds. The understanding that these these resonances are bound or virtual states made of the corresponding heavy meson pair is strongly supported by an analysis of the behavior of heavy quark spin in production and decay of these resonances~\cite{bgmmv}. Namely the strength of the interaction depending on the spin of a heavy quark in QCD is inversely proportional to the quark mass. For this reason there is an approximate Heavy Quark Spin Symmetry (HQSS) for heavy quarks which symmetry is exact in the limit of infinite quark mass, and is a very good approximation for soft processes involving bottom quarks. An illustration of the quality of this approximation can be found e.g. in the relative rate of HQSS suppressed and allowed transitions between the bottomonium states:  $\Gamma[\Upsilon(2S) \to \Upsilon(1S) \eta]/\Gamma[\Upsilon(2S) \to \Upsilon(1S) \pi \pi] \sim 10^{-3}$. In a widely separated meson-antimeson pair the spin of the $b$ quark (antiquark)  is fully correlated with the spin of the light antiquark (quark) that is contained in the corresponding pseudoscalar or vector $\bar B$ ($B$) meson. For this reason the spins of the $b$ and $\bar b$ are not correlated with each other and the $b \bar b$ pair is in a mixed spin state:
\bea
&&B^* \bar B - \bar B^* B \sim 0^-_H \otimes 1^-_L + 1^-_H \otimes 0^-_L \nonumber \\ &&B^* \bar B^* \sim 0^-_H \otimes 1^-_L - 1^-_H \otimes 0^-_L~,
\label{hls}
\eea
where $S_H$ ($S_L$) stands for the total spin of the heavy (light) quark-antiquark pair. If the $Z_b$ resonances contain widely separated heavy meson-antimeson pairs, one would expect that the spin structures (\ref{hls}) should be to some accuracy retained in the two molecular states. It is exactly this behavior that has been observed in experiment~\cite{bellezb} with the $Z_b$ resonances decaying with comparable rates to the ortho- ($S_H=1$) and the para- ($S_H=0$) states of bottomonium with emission of a pion: $Z_b \to \Upsilon(nS) \pi$ and $Z_b \to h_b(kP) \pi$. Moreover the relative signs of the transition amplitudes implied by this picture~\cite{bgmmv} are in a remarkable agreement with the observed relative phase of the two $Z_b$ resonance contribution to the processes $\Upsilon(5S) \to Z_b \pi \to \Upsilon(nS) \pi \pi$ and $\Upsilon(5S) \to Z_b \pi \to h_b(kP) \pi \pi$.

The spin structure (\ref{hls}) of free meson pairs is  preserved withing the $Z_b$ resonances inasmuch as the dependence of the interaction through light degrees of freedom on the light spin $S_L$ is not essential. In particular, 
such dependence would induce transitions between the two states of meson pairs in Eq.(\ref{hls}) and give rise to decay  of the heavier $Z_b$ resonance  to the lighter meson pair, $Z_b(10650) \to B^* \bar B + $c.c., which decay is perfectly allowed and can only be suppressed due to the spin orthogonality in Eq.(\ref{hls}). The data~\cite{bellezbb} show no indication of such decay, so that possibly the interaction within the heavy meson-antimeson pairs indeed does not depend on the spin of light quarks implying an existence of a certain `accidental' Light Quark Spin Symmetry (LQSS). Such symmetry is not expected within QCD. Moreover it is explicitly broken by pion exchange. The observed suppression of a spin-dependent interaction then can be described in terms of a form factor suppression~\cite{mv16} of the pion exchange or as an effect of a contact term~\cite{wangetal} effectively canceling the pion contribution. 

If the light-quark interaction between the heavy meson can indeed be approximated as spin-idependent, one should then expect~\cite{mv11} existence of near-threshold resonances in the $S$-wave states of the meson pairs related to the $Z_b(10610)$ and $Z_b(10650)$ by the quark spin symmetry. There are four such states:
\begin{widetext}
\bea
\label{wbs}
&&W_{b2}: ~~1^-(2^+):~~ \left. \left ( 1^-_H \otimes 1^-_{L} \right ) \right |_{J=2}~, ~~~ B^* \bar B^*~; \\ \nonumber
&&W_{b1}: ~~1^-(1^+):~~ \left. \left ( 1^-_H \otimes 1^-_{L} \right ) \right |_{J=1}~, ~~~ B^* \bar B + \bar B^* B;
\\ \nonumber
&&W_{b0}': ~~ 1^-(0^+):~~ {\sqrt{3} \over 2} \, \left ( 0^-_H \otimes 0^-_{L} \right ) + {1 \over 2} \,\left. \left ( 1^-_H \otimes 1^-_{L} \right ) \right |_{J=0}~, ~~~ B^* \bar B^* ~;
\\ \nonumber
&&W_{b0}: ~~1^-(0^+):~~ {1 \over 2} \, \left ( 0^-_H \otimes 0^-_{L} \right ) - {\sqrt{3} \over 2} \,\left. \left ( 1^-_H \otimes 1^-_{L} \right ) \right |_{J=0}~, ~~~  B \bar B~,
\eea
\end{widetext}
where the quantum numbers $I^G(J^P)$ and the corresponding meson-antimeson thresholds are indicated. Due to their negative $G$ parity the $W_{bJ}$ states cannot be produced by a single pion emission from a bottomonium-like resonance produced in $e^+e^-$ annihilation. They can be produced by emission of a photon in similar processes, however the rate is possibly very small. The lowest mass resonance $W_{b0}$ may be accessible in a two pion process $\Upsilon(6S) \to W_{b0} \pi \pi$, however the rate of this process is hard to predict. Finally, the most favorable setting for a search in the process $e^+e^- \to W_{bJ} \rho$  requires c.m. energy in excess of 11.4 - 11.5\,GeV.

An application of the same considerations to charmonium-like molecules made of charmed mesons has some peculiarity due to apparently weaker constraints from the HQSS and also due to a significant isotopic breaking near threshold arising from the  mass difference between the neutral and charged $D^{(*)}$ mesons. For instance, the $X(3872)$ peak has the $J^{PC}$ quantum numbers similar to the neutral component of the $W_{b1}$ isotopic triplet. However it is not clear what is the significance of the strong isospin breaking in the masses of the charmed mesons for the emergence of the $D^{*0} \bar D^0$ threshold peak. For this reason it would be troublesome to predict a full structure of $S$ wave threshold molecules made of charmed mesons. The $Z_c(3900)$ and $Z_c(4020)$ appear to be hidden-charm analogs of the bottomonium-like $Z_b$ resonances. However the heavy charmed quark spin properties are not as clear cut. The $Z_c(3900)$ resonance has been observed in the channel $J/\psi \pi$~\cite{besiiiz39} but not (yet?) in $h_c \pi$~\cite{besiiiz40}.  For the peak $Z_c(4020)$ the situation is reversed: the decays to $h_c \pi$ are `seen' but those to $J/\psi \pi$ are `not seen'. It thus appears that the mixing of the ortho- and para- spin states of the $c \bar c$ pair is less straightforward in this case than for the $Z_b$ resonances although some hints at decays of $Z_c(3900)$ to the paracharmonium states, $h_ \pi$ and $\eta_c \rho$ are reported~\cite{yuan}. Furthermore, the heavier resonance $Z_c(4020)$ appears not to decay into the lighter meson pair $D^* \bar D +$c.c. much in the same manner as its bottomonium-like counterpart $Z_b(10650)$. 

\section{Hadro-quarqonium}

Some manifestly exotic hidden-charm resonances have masses that are not particularly close to any two-body thresholds, so that it would be troublesome to interpret them as molecular states. These states include $Z(4430)^\pm$~\cite{pdg} decaying to $\psi(2S) \pi^\pm$, $Z_c(4100)^\pm$~\cite{lhcbzc41} and $Z_c(4200)^\pm$~\cite{bellezc42} in the respective channels $\eta_c \pi^\pm$ and $J/\psi \pi^\pm$, and a pair of resonances $Z_c(4050)^\pm$ and $Z_c(4250)^\pm$ observed~\cite{bellezc2} in the decay channel $\chi_{c1} \pi^\pm$. (A subsequent search~\cite{babarzc2} for the latter two peaks however turned to be unsuccessful, neither any other confirmation was reported in over a decade,  so that the experimental status of these two states is not quite clear.) The decay channels with charmonium are essentially the only experimentally observed decay modes, while less, or very little, is known about decays of these peaks into final states with open charm. This implies that the latter states do not entirely dominate the decay modes of the exotic hidden-charmed resonances. (As an example of such saturation of the width by open charm channels the resonance $\psi(3770)$ can be mentioned.) 

Such behavior of the exotic resonances strongly suggests~\cite{mvch,dv} that their structure is dominantly described by the hadro-charmonium picture shown as the second type of configuration in the Figure~\ref{inside}. The heavy $c \bar c$ pair is a state of charmonium embedded in an excited light-quark matter due to a two (or multi-) gluon interaction. The binding of charmonium in light-quark matter has been considered long ago~\cite{bst,kv,sv} in terms of interaction of charmonium inside nuclei. The hadro-charmonium resonances are different in that instead of the nucleus the compact $c \bar c$ state is inside a spatially large light-matter excitation. In this picture the decays into charmonium and light mesons are due to a de-excitation of the light matter with the $c \bar c$ state remaining. The decay width is then set by the scale typical for excited light-quark resonances and should thus be broad - in tens or hudreds of MeV. It is thus expected that such decays mostly proceed into the charmonium state that is already contained `inside', in agreement with the observed appearance of particular $c \bar c$ in the decay products (e.g. $\psi(2S)$ rather than $J/\psi$ in the decays of  $\psi(4330)^\pm$). Naturally, one can expect that `other' $c \bar c$ states should be present among the decay products at a sub-dominant level due to a `deformation' of charmonium by the binding.

\begin{figure}[h]
\centering
\includegraphics[width=80mm]{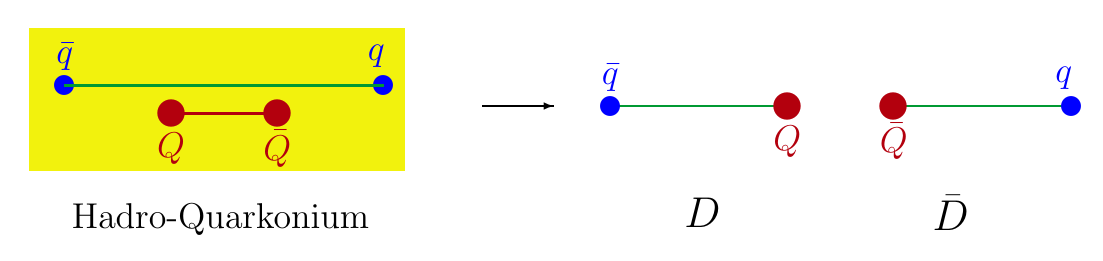}
\caption{Reconnection of binding in hadro-charmonium decay into open heavy flavor} \label{hqdecay}
\end{figure}

\begin{figure}[h]
\centering
\includegraphics[width=80mm]{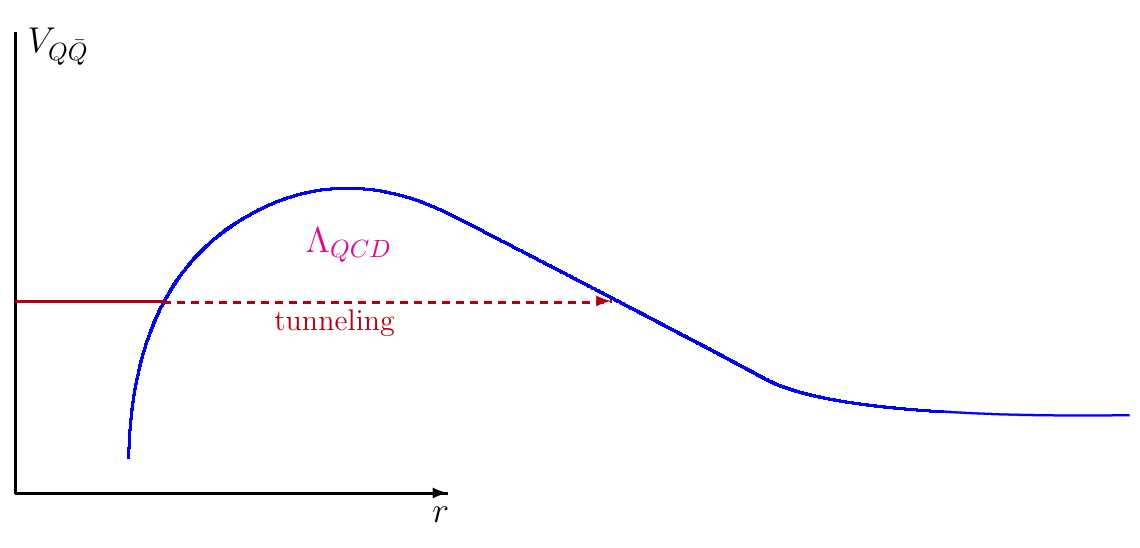}
\caption{Behavior of a Born-Oppenheimer type potential between heavy quarks} \label{hqpot}
\end{figure}

The tendency of hadro-quarkonium to not decay overwhelmingly into states with open heavy flavor can be semi-quantitatively argued in the limit of large heavy quark mass~\cite{dgv}. Indeed, such decay requires a reconnection of the dominant bindings within the four-quark complex, as illustrated in Figure~\ref{hqdecay}.
In the limit of large heavy quark mass $M_Q$ one can consider an effective potential between the heavy quark and antiquark in terms of Born-Oppenheimer approximation, where the potential is  the energy of the system as a function of distance between static $Q$ and $\bar Q$. Clearly, at short distances there is a Coulomb-like potential well, and at long distances the energy goes to a constant corresponding to widely separated heavy mesons. At intermediate distances however, the energy is increased due to the gluonic bindings being out of the minimal energy state, as shown in Figure~\ref{hqpot}.
The discussed decay into open heavy flavor channel can then be considered as tunneling through the potential barrier. Given that the parameters of the potential are determined by $\Lambda_{QCD}$ one can expect the dependence of the tunneling rate on the mass $M_Q$ as
\be
\Gamma({\rm open~flavor}) \sim \exp \left ( - C \, \sqrt{M_Q/\Lambda_{QCD}} \right ) ~,
\label{gtun}
\ee
where $C$ is a numerical constant.

The dominant QCD interaction giving rise to the analog of van der Waals force between a compact quarkonium and light matter is the chromoelectric dipole $E1$ that does not depend on the heavy quark spin. Thus in the hadro-charmonium picture one can expect that the exotic states appear in multiplets whose components are related by HQSS in the same way as the states of quarkonium are related by this symmetry. It is very likely that an example of such multiplet is provided by the $Z_c(4100)$ resonance decaying into $\eta_c \pi$ and the $Z_c(4200)$ observed in the $J/\psi \pi^\pm$ channel. The quantum numbers $J^P$ of these resonances are not yet well known, but the data are quite compatible with them being $0^+$ and $1^+$ respectively. One can thus suggest~\cite{mvzc} that these are hadro-charmonium resonances made by embedding in the $S$ wave the $\eta_c$ or $J/\psi$ charmonium into the same excited pion-like hadronic state. Clearly the mass difference between the resonances agrees (within expeted accuracy) with that between $J/\psi$ and $\eta_c$, and the observed total widths do not contradict within the errors to the expectation that they should be equal. Furthermore, the rate of production in $B$ decays (also within errors) agrees with the expected relation
\be
{{\cal B}[B^0 \to Z_c(4100)^- K^+] \over {\cal B}[B^0 \to Z_c(4200)^- K^+]} \approx \left . {{\cal B}[B^0 \to \eta_c \pi^- K^+] \over {\cal B}[B^0 \to J/\psi \pi^- K^+]} \right |~.
\label{simy}
\ee
The leading HQSS breaking interaction in QCD is well known, so that one can also predict a relation between sub-dominant heavy-spin violating decays~\cite{mvzc}
\be
\Gamma[Z_c(4100) \to J/\psi \rho] \approx 3 \, \Gamma[Z_c(4200) \to \eta_c \rho]~,
\label{rzcr}
\ee
with the branching fraction for each of these processes expected in the ballpark from several percent to a few tens percent relative to the observed decays to $\eta_c \pi$ and $J/\psi \pi$.

If the $Z_c(4100)$ and $Z_c(4200)$ are identified as the $1S$ states of charmonium `stuck' in an excited pion, then it is possible that there are similar states with an excited Kaon instead of a pion, that are about 150\,MeV heavier: $Z_{cs}(4250)$ and $Z_{cs}(4350)$~\cite{mvzcs}. These strange hadro-charmonia should be SU(3) flavor symmetry partners of the non-strange ones, in the same manner as the excited resonances $K(1460)$ and $\pi(1300)$ are partners. It scan be expected that the flavor SU(3) symmetry should be applicable to hadro-quarkonium with about the same accuracy as to ordinary light hadrons. This expected behavior is quite different from that for molecular states. Indeed, in the latter systems the SU(3) breaking by the strange quark mass is a large effect in comparison with the binding/excitation energy of a molecular state. Even the isotopic mass differences can be of importance in the molecules, as is the case for $X(3872)$. The strange hadro-charmonium resonances, decaying to $\eta_c K$ and $J/\psi K$ should be observable in the decays of the strange $B_s$ mesons, $B_s \to Z_{cs} K$, at the same rate~\cite{mvzcs} (corresponding to the branching fraction $\sim 10^{-5}$) as the non-strange ones, $B \to Z_c K$.  

\section{Hidden-charm pentaquarks}

Some time ago the LHCb experiment has reported~\cite{lhcbpc1,lhcbpc2} an observation of resonant structures in the hidden-charm pentaquark channel $J/\psi p$ produced in decays of the $b$ hyperon $\Lambda_b \to J/\psi p K^-$. The initial observation indicated a broad, $\Gamma = 205 \pm 18 \pm 86\,$MeV, structure $P_c(4380)$ and a narrower peak  $P_c(4450)$ with $\Gamma= 39 \pm 5 \pm 19\,$MeV. Very recently a refined picture was presented~\cite{lhcbpc3} with the peak near 4450\,MeV resolved into two narrower ones, $P_c(4440)$ with $\Gamma = 20.6 \pm 4.9^{+8.7}_{-10.1}\,$MeV and $Pc(4457)$ with $\Gamma = 6.4 \pm 2.0^{+5.7}_{-1.9}\,$MeV, and an additional observed narrow peak $P_c(4312)$ with $\Gamma = 9.8 \pm 2.7^{+3.7}_{-4.5}\,$MeV. 

The models of the internal dynamics of the barionic pentaquark states essentially follow the same lines as in the previous discussion of the mesonic four-quark systems. In particular, the resonances with mass in the vicinity of a threshold for a charmed hyperon and a charmed (anti)meson can be tested for being of the molecular kind. The newly reported three narrow resonances tantalizingly suggest~\cite{lhcbpc3} such a structure composed of $\Sigma_c$ and $\bar D^{(*)}$ bound in the $S$-wave. Namely, the thresholds for $\Sigma_c^+ \bar D^0$ and $\Sigma_c^+ \bar D^{*0}$ are higher than the central values of the masses of $P_c(4312)$ and $P_c(4457)$ by respectively about 5 and 2 MeV, while the $P_c(4440)$ could have binding energy of about 20 MeV due to the spin-spin interaction of $\Sigma_c$ and $\bar D^*$ through the light degrees of freedom. In this case the quantum numbers $J^P$ of the three states should be $1/2^-$, $1/2^-$ and $3/2^-$, which assignment does not contradict the data, but is not yet established either. The isotopic spin of a baryonic hidden-charm state $X_{c \bar c}$ produced in the decays $\Lambda_b \to X_{c \bar c}^+ K^-$ due to the underlying quark process $b \to c \bar c s$ is necessarily equal to 1/2. However the isospin violation by the mass differences in the $\Sigma_c$ and $D^{(*)}$ isotopic multiplets  can be enhanced in the pentaquarks due to closeness of the threshold~\cite{guoetalpc}.  In the  molecular model one should expect a strong decay of pentaquarks into the final state(s) with $\Lambda_c$ instead of $\Sigma_c$: $P_c \to  \Lambda_c \bar D^{(*)}$. Indeed, there is approximately 160 - 300 MeV of energy available for such decays, and there appears to be no principle forbidding them to proceed from a molecular state due to the $\Sigma_c \bar D^{(*)} \to \Lambda_c \bar D^{(*)}$ scattering. The experimental status of these decays is not clear at present. However if they are not found, the molecular model may have difficulty explaining their suppression. An alternative model for pentaquarks , based on the hadro-quarkonium picture and free from this potential difficulty is developed in Refs.~\cite{epp1,ep,epp2}.

Further studies of hidden-charm pentaquarks would be greatly facilitated if additional to the production at LHC ways of experimentation with them could be found. One such suggested~\cite{wlz,kuvo,kr} alternative source of pentaquarks decaying to $J/\psi p$ is their formation in the $s$ channel by photon beam on hydrogen target, $\gamma p \to P_c$. First data from the GlueX experiment searching for this process have just appeared~\cite{gluex} with no evidence yet of  pentaquark resonance(s), and setting a model-dependent upper limit of approximately 2\% on the branching fraction ${\cal B}(P_c \to J/\psi p)$. Given that this is just a first measurement and the study is currently in flux, it appears to be premature to draw from the reported data any far reaching conclusions on the existence and the properties of the pentaquarks.

\begin{figure}[h]
\centering
\includegraphics[width=80mm]{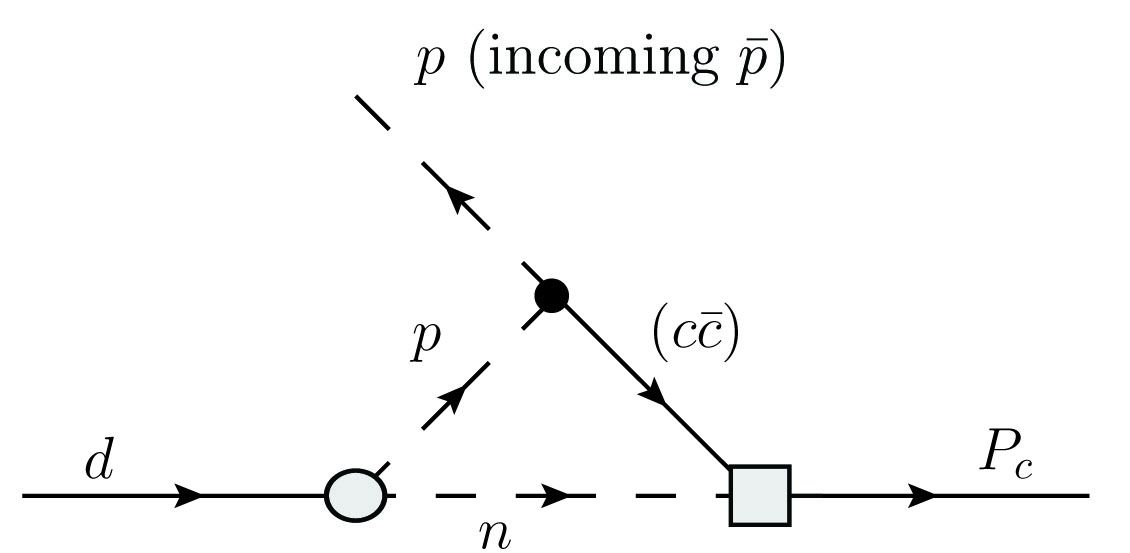}
\caption{The graph for the hidden-charm pentaquark formation in $\bar p - d$ collision. Dashed lines denote nucleons, the solid lines, as marked, are for the deuteron ($d$), the charmonium state $(c \bar c)$ and the pentaquark $P_c$.} \label{pentad}
\end{figure}

Another possible source of hidden-charm pentaquarks can be provided by antiproton - deuterium collisions~\cite{mv19} and can be studied e.g. in the PANDA experiment~\cite{panda} at FAIR. The mechanism of formation in the $s$ channel of a resonance coupled to a charmonium state and a nucleon $P_c \to (c \bar c) + N$ is shown in Figure~\ref{pentad}.
 
The dominant part of the wave function of the nucleons inside the deuteron, treated as a loose bound state, can be effective only if the kinematical constraints in the graph do not require the relative momentum of the neutron and the proton in the triangle to be large in comparison with the inverse nucleon size. Considering the process in the rest frame of the deutron (which frame coincides with the lab frame in a realistic experiment, e.g. in PANDA), one readily finds that both nucleons in the triangle can be on-shell and simultaneously at rest if the mass $M$ of the pentaquark is related to the mass $m$ of the charmonium state and the nucleon mass $\mu$ as $M=M_0(m)$ with 
\be
M_0^2(m) = 2m^2 + \mu^2~.
\label{m0}
\ee
(The small binding energy $\epsilon = -2.22\,$MeV in the deuteron is obviously neglected in this expression.) In particular, for the $(c \bar c)$ charmonium mass of $J/\psi$ and $\eta_c$ the special value of the pentaquark mass is estimated as respectively $M_0(m_{J/\psi}) = 4.48\,$GeV and $M_0(m_{\eta_c}) = 4.33\,$GeV. It can be readily noted that the former of these values is quite close to the measured mass of $P_c(4440)$ and $Pc(4457)$, while the latter is close to the mass of $P_c(4312)$. It can be noted in connection with this kinematical observation that unlike mesonic states the same pentaquark resonance can generally couple to both $J/\psi N$ and $\eta_c N$channels. The low-momentum wave function for the motion inside the deuteron is applicable in a range of the pentaquark mass around the special value (\ref{m0}) so that in fact all the so far reported $P_c$ states can be studied in both decay channels. The resulting expected cross section at the maximum of the Breit-Wigner peak for a pentaquark can be estimated in terms of the branching fraction ${\cal B}[P_c \to (c \bar c) + n]$ as~\cite{mv19}
\bea
&&\sigma (\bar p + d \to P_c) \sim  \nonumber \\
&&10^{-33}\, {\rm cm}^2 \, \left \{ {\Gamma[(c \bar c) \to p \bar p] \over 1\,{\rm keV} } \right \} \, {\cal B}[P_c \to (c \bar c) + n]~,
\label{spc}
\eea
This estimate clearly appears to favor studies in the $\eta_c+n$ channel due to a much larger than for $J/\psi$ decay width into $p \bar p$: $\Gamma(\eta_c \to p \bar p) \sim 50\,$keV.
 
\section{Remarks on Di-diquarks}

It is mentioned in the introduction that the third type of configuration in Figure~\ref{inside} with dominant correlations being within diquarks does not have a justification within QCD. Here I would like to somewhat expand on this remark. The usual argument in favor of dominant configurations with color antisymmetric diquark pairs is that there is an attraction in the antisymmetric state and a repulsion in the symmetric one (a detailed discussion of the di-diquark model can be found in the review~\cite{epp}). This argument is based on considering a one-gluon exchange between the quarks. Following this argumentation and assuming that the one-gluon exchange provides a relevant guidance, it is helpful to consider in full the one-gluon exchange potential in a system of two quarks and two antiquarks constrained by the condition that the system is an overall color singlet. Denote the constituents of the system as $q_1 \bar q_2 q_3 \bar q_4$, where the subscripts stand for the variables, e.g. the coordinates $\vec r_i$, of the quarks and the antiquarks (the odd are for the quarks and the even for the antiquarks). There are two orthogonal color configurations corresponding to the overall color neutrality: with color-symmetric diquarks $u = \{q_1 q_3\} \{\bar q_2 \bar q_4\}$ and with color-antisymmetric $w=[q_1 q_3] [\bar q_2 \bar q_4]$, where the curly (straight) braces stand for the color symmetrization (antisymmetrization). The one-gluon exchange in fact mixes these two configurations, so that the potential $V$ in the space of $(w,u)$ has the form of a matrix~\cite{clv} written in terms of the Coulomb factors $c_{ij} = \alpha_s/|\vec r_i - \vec r_j|$:
\be
V =  - {1 \over 4} \, \left ( \begin{array}{cc} {N_c^2 -1 \over N_c} \, r + {N_c+1 \over N_c} \, t & \sqrt{N_c^2-1} \, s \\ \sqrt{N_c^2-1} \, s & {N_c^2 -1 \over N_c} \, r - {N_c - 1 \over N_c} \, t 
\end{array}
\right ) \, ,
\ee
with $N_c$ being the number of colors and the notations are used: $r= c_{12}+c_{34}+c_{14}+c_{23}$, $s=c_{12}+c_{34}-c_{14}-c_{23}$, $t= 2 c_{13}+2 c_{24} - c_{12}-c_{14}-c_{23}-c_{34}$.
The attraction (repulsion) within the antisymmetric (symmetric) diquark is described by the term $t$ while the term $s$ describes the mixing between these configurations. One can readily see that at large number of colors the $t$ term is by the factor $N_c^{-1}$ smaller than the $s$ term (and these terms are comparable at $N_c=3$). Thus the mixing is parametrically more important at large $N_c$ than the difference between the attraction and repulsion  (or at least equally important at $N_c=3$) and the configurations with either specific symmetry cannot be dominant. This conclusion applies if there are no other parameters that would compensate for the color suppression. The only situation where such `overriding' parameter is present is in a system with a double-heavy diquark, $QQ\bar q \bar q$, where the quarks $Q$ are very heavy in comparison with either the masses of the antiquarks $\bar q$, or with $\Lambda_{QCD}$ if the antiquarks are light. In this case the Coulomb attraction within the color antisymmetric $QQ$ pair is sufficiently enhanced by the mass of $Q$ in order to overcome the mixing~\cite{clv}. In the hidden-charm or hidden-bottom multiquark systems there is no such parameter and there is no grounds whatsoever to consider the model with dominantly color-antisymmetric diquarks.

\begin{acknowledgments}
This work is supported in part by U.S. Department of Energy Grant No.\ DE-SC0011842.
\end{acknowledgments}

\bigskip % extra skip inserted

\end{document}